\documentclass[conference]{IEEEtran}
\usepackage[utf8]{inputenc}
\usepackage{graphicx}
\usepackage{hyperref}

\title{MIGRATION FROM RURAL TO URBAN: CASE ANALYSIS OF THE PEASANT COMMUNITY OF ANTACAHUA, PUNO-2023}

\author{
  \IEEEauthorblockN{Laymir Sebastian Apaza Ajrota}
  \IEEEauthorblockA{Universidad Nacional del Altiplano Puno\\
  \texttt{laapazaa@est.unap.edu.pe}}
  \vspace{5mm}

  \IEEEauthorblockN{Clever Joel Mamani Ari}
  \IEEEauthorblockA{Universidad Nacional del Altiplano Puno\\
  \texttt{clmamaniar@est.unap.edu.pe}}
  \and
  \IEEEauthorblockN{William Yoel Incacutipa Incacutipa}
  \IEEEauthorblockA{Universidad Nacional del Altiplano Puno\\
  \texttt{wincacutipa@est.unap.edu.pe}}
  \vspace{5mm}
   \IEEEauthorblockN{Renzo Apaza Cutipa}
  \IEEEauthorblockA{Universidad Nacional del Altiplano Puno\\
  \texttt{renzo@unap.edu.pe}}
}
\begin{document}
\maketitle

\begin{center}
    \textbf{Abstract}
\end{center}

This research was conducted based on the observation of the low presence of people in the Community Campesina de Antacahua. The aim was to identify the reasons influencing the migration of the community and determine if it is related to rural-to-urban migration trends. Additionally, normative alternatives are proposed to combat this phenomenon. The research employed a mixed approach with a descriptive-correlational scope, conducted through a survey of community members, as well as normative and bibliographic analysis.

The results demonstrate that the emigration from the Community Campesina de Antacahua is driven by factors such as dissatisfaction with the lack of basic services such as water, electricity, internet, telephone, education, and others (34.4\%), lack of economic support (23.4\%), low crop productivity (15.6\%), reasons related to studies (18.8\%), and lack of land (7.8\%). According to the age group of emigrants, 12.5\% are teenagers (15–22 years old), 17.2\% are young adults (23–29 years old), 43.7\% are young adults (30–36 years old), and 26.6\% are adults (37–60 years old). The preferred migration destinations are the city of Juliaca (40.0\%), the city of Puno (23.0\%), mines located in Ananea and Cojata (25.0\%), and other places (12.0\%).

Literature analysis revealed that government-implemented public policies are insufficient to improve the quality of life for community members. To prevent migration from the Community of Antacahua, it is recommended to promote the installation of basic services by providing land to entities or companies willing to offer services. Additionally, fostering economic activity and employability could be achieved by incorporating a new member into the Communal Directive.

\section{INTRODUCTION}
Migration is a phenomenon that is present in our daily lives. As society progresses, people migrate from their places of origin to their destinations with the intention of finding better job opportunities that allow them to improve the quality of life for their families (Gutiérrez Silva et al., 2020). According to León (2015) and Guillén et al. (2019), migration is caused by social, economic, and political problems, leading to an increase in inequality, which constitutes a significant aspect of discrimination within and between social classes. This is attributed to poor government policy management. Lazos Chavero (2020) warns that the lack of young people in rural areas, the absence of integration policies, and limited support for small producers, among other factors, are the main threats in the Mexican countryside.
In the Antacahua Peasant Community, which is predominantly rural, it was observed that there is a migrant population, both permanent and temporary, resulting in a reduced presence of young people. This research describes the factors influencing the youth population to migrate from the Antacahua Peasant Community. It also analyzes normative policies that can be incorporated into the Statute of the Antacahua Community, within the framework of the organizational autonomy granted by the Political Constitution of Peru and the General Law of Peasant Communities, to make migration less attractive.
The objectives of this research were: to describe the factors influencing the residents of the Antacahua Community to migrate; to analyze the normative framework of government public policies to combat rural abandonment; and to propose normative principles that can be included in the community's statute to make migration less attractive for its population.
Undoubtedly, the problem of low population in all age groups brings consequences such as land unproductivity, family abandonment, declining local economy, among others. Identifying the main reasons allowed proposing alternatives to combat or reduce migration. Understanding the legal aspects of government public policies and relating them to the main problem, which is the migration of the population in the Antacahua community, allowed proposing solutions or alternatives at the level of normative provisions that have an impact on the community's organization, as Peasant Communities are legal entities of public law, autonomous in their organization.

\section{MIGRATION FROM RURAL TO URBAN AREAS}

In Peru, according to the 2017 census conducted by the National Institute of Statistics and Informatics (INEI), the urban population represents 79.3\%, while the rural population represents 20.7\%. These figures have increased and decreased, respectively. The urban population grew by 17.3\% between 2007 and 2017, with an average annual rate of +1.6\%. Conversely, the rural population decreased by 19.4\% during the same period, at an average annual rate of -2.1\%. As evident, the rural population in Peru is declining, indicating migration from rural to urban areas.
The 2017 census also reports a growth in the population of the Coastal region (13.8\%) and the Jungle region (10.9\%) over a 10-year period (2007-2017), demonstrating greater geographical dynamism. In contrast, the Sierra region experienced a negative variation (-5.7\%). This information suggests that especially the Sierra region is not attractive for receiving migratory movements or establishing permanent residences.
According to the IV National Agricultural Census 2012, the Sierra region of Peru hosts the largest number of Peasant Communities, both in terms of quantity and area. Notable examples include Puno with 1388 (22.7\% of territory), Cusco with 977 (16.0\% of territory), Huancavelica with 617 (10.1\% of territory), Ayacucho with 560 (9.2\% of territory), and others. It's worth mentioning that the Sierra is 60\% rural, and the Rural Sierra has the highest total poverty percentage in the country at 37.8\% (INEI, 2017). This indicates a high poverty rate in the Rural Sierra, where most Peasant Communities are situated. The I Census of Peasant Communities 2017 shows that out of the 6,682 censused communities, 69.0\% had a population that migrated in the last 12 months.
The predominant migratory flow is from rural to urban areas, and from the Sierra to the Coastal or Jungle regions, indicating a population loss for Peasant Communities in the Sierra and rural areas. Determining factors for these migratory movements include economic growth and decentralization. Decentralization helps solidify the workforce in regions and is closely tied to the recovery of agricultural activities. The dynamics of mining in the Sierra are related to the demand for services such as construction, transportation, food, health, and education. Poverty is a significant factor, especially in the Rural Sierra, with a poverty rate of 61.2\% (INEI, 2011), prompting migration to cities. Government-implemented employment opportunity policies are gradually reducing poverty in the Rural Sierra. The economically active population in the Sierra is primarily engaged in agriculture, often facing poverty or extreme poverty (INEI, 2011), indicating that agriculture is not a profitable activity. Education plays a role, as adolescents migrate in search of new opportunities, and youth in the rural Sierra who receive higher or technical education rarely, if at all, return to work in rural areas. Basic services are crucial factors determining the comfort of rural dwellings; thus, a rural dwelling without basic services is not attractive, leading to migration from rural to urban areas.

\section{CONSEQUENCES OF YOUTH MIGRATION}

The phenomenon of migration has a higher incidence among young individuals, either due to family or personal decisions seeking a better quality of life in urban areas. This trend is observed globally, and in our country, there has been significant internal migration since the late 20th century, leading to various effects on urban areas and the peasant communities from which inhabitants are moving, resulting in the Coastal region of Peru having a higher population rate (Aruj, 2008).

\subsection{Demographic Effects:}

The demographic effects of migration are factors that impact natural and structural growth within a population losing inhabitants. This can lead to the disappearance of rural populations in a relatively short period, causing an increase in birth rates and a decrease in mortality rates in favored regions for migrants. However, the influx of migrants introduces another issue: the competition for opportunities to achieve a decent and comfortable quality of life in their new environment (INEI, 2020).
\subsection{Economic Effects:}

There are various economic consequences, with the most notable being

\subsubsection{Employment and Salaries: }

A negative effect of migration is an elevated unemployment rate, and native workers may experience lower wages or tend to accept more strenuous jobs to avoid displacement. Immigrants increase the labor supply, competing for jobs with those in strenuous occupations, who may be exploited due to fear of job loss. Some of these workers may be displaced and see reduced wages. However, data from various contexts indicate that the eventual decline in native workers' wages, attributable to the labor supply of immigrants in certain formal economic sectors, is essentially trivial or non-existent, particularly in informal sectors (INEI, 2015).

\subsubsection{Economic Growth and Productivity: }

Migration has a positive effect on the economic growth of receiving regions, as migrants contribute to increasing production by expanding the available workforce. The human capital of migrants, acquired through work experience in their places of origin, is leveraged. Migrants can also boost productivity by enabling native workers to move from poorly paid jobs with low benefits to higher-skilled and better-paid positions, increasing production possibilities and, consequently, economic growth (CONAPO, 2012, p. 25).

\subsubsection{Fiscal Costs and Public Services: }

A problematic aspect at the state level involves assessing the costs and benefits of migration in destination regions, considering the costs migrants represent for the state and public assistance programs. The argument is that the arrival of immigrants and their families, many of whom seek health and education benefits for their children, increases the costs of the social system and creates imbalances in government accounts. This is often a contentious debate in receiving societies and political elites. However, most of the population in the receiving region sees maintaining harmony without considering immigrants as more important. This occasionally translates into the enactment of laws seeking to limit immigrants and their descendants' access to education, health, and other social protection services. However, this is not clearly observed in Peru, as the potential risks of increased migration from urban areas have not been prominently raised (CONAPO, 2012, p. 25-26).

\subsection{Social Effects:}

The most significant factor is termed integration, reflecting the challenges and difficulties at the sociocultural and economic levels, particularly regarding cultural differences observed among new urban inhabitants. Despite attempts by newcomers to assimilate and integrate into the lifestyle of their new region, the traits of their original culture seldom disappear, maintaining the challenge of cultural diversity and syncretism. Additionally, as rural areas have a globalized culture, this leads to cultural loss among migrants (CONAPO, 2012, p. 26).
Another negative factor in rural communities arises when migration is highly significant, as the loss of population can diminish the productive potential of the origin communities. The temporary or permanent departure of individuals results in a shortage of labor in certain sectors, such as agriculture and livestock, thereby discouraging economic growth (INEI, 2020).
Within the aforementioned consequences, it is essential to consider existing inclusion gaps. As a positive measure to address these issues, national programs have been implemented to integrate migrants into urban areas. Some notable national programs include "QaliWarma" for food and nutrition, "Cuna Más" and "Sistema Integral de Salud" for health and well-being, and "Jóvenes Productivos" and "Pensión 65" for economic support and productivity. These social programs and initiatives, provided by the State to individuals in vulnerable and impoverished situations in Peru, are promoted by the Ministry of Development and Social Inclusion (MIDIS). They contribute to social integration and the sustainability of migrations, benefiting both urban and rural areas (Abramo, Cecchini, y Morales, 2019).

\section{OPPORTUNITIES FOR YOUTH IN PEASANT COMMUNITIES AND URBAN AREAS}

When considering young individuals from rural or urban areas, both cases present opportunities influenced by various economic, social, or educational factors. The youth population seeks a better quality of life, compelling them to engage in any type of work that can generate income. Individuals from rural or urban areas have different opportunities depending on the skills they possess or have acquired.
According to Weller, internal migration among the youth is a symptom of modern times. Highly mobile and low-quality job markets, without clear references, contribute to uprooting the workforce. Additionally, factors such as globalization, accessible communication, and information technologies in the modern world compel young people to leave their rural homes and migrate to urban centers, both within and outside their country, often in search of new job and economic opportunities. This process acts as a factor that increases dynamism in the job market (Weller, 2006).

\subsection{Opportunities in Rural Communities: }

Development opportunities in rural areas are currently often scarce due to the gradual impact of the migratory process on employability models. However, the country depends on certain agricultural and livestock producers, as well as producers in the food industry, such as fish farmers, where employment opportunities will always exist (Espejo, 2017).

\subsection{Opportunities in Urban Areas: }

In cities within more developed departments, young individuals entering the workforce are characterized by low wages, job insecurity, and discrimination based on race and gender. Additionally, employers seek young people with good educational backgrounds, a situation that does not favor rural areas. This highlights concerning gaps between the opportunities available to young people living in urban and rural areas (Espejo, 2017).

\section{RETURN MIGRATION}

When we think about migration, we often have the idea of a linear situation where individuals or groups emigrate and permanently settle in the "new" place. However, contemporary migration is increasingly multidirectional, and the return to countries of origin is evident, driven by various causes.
In research by the Global Migration Data Portal (2022), return migration is explicitly defined when a migrant returns to their point of origin. This can occur within an international context, between the territories of different countries, or within the borders of a single country.

\subsection{Types of Return Migration:}

According to the Glossary of the International Organization for Migration (IOM, 2022), there are four types of return migration: spontaneous return, forced return, assisted return, and voluntary return.

\subsubsection{Spontaneous Return: }

Spontaneous return occurs when a migrant or a group of migrants abruptly returns to their country or place of origin. In general, migrants do not receive any support or assistance, either national or international, during their return journey.

\subsubsection{Forced Return:}

Forced return happens when migrants are compelled to return to their places of origin or any other location that accepts them. This type of return usually occurs when there are legal or administrative issues, leading to the "return" of migrants against their will.

\subsubsection{Voluntary Return: }
Voluntary return is an independent decision made without assistance from any entity. It is based on the voluntary choice of individuals who wish to return to their place of origin or another destination.

\subsubsection{Assisted Return: }
Assisted return is closely linked to voluntary return, but migrants receive administrative, logistical, and often financial support. Assistance is provided to help them initiate the reintegration process in their country of origin. For this type of return, two requirements are considered: a) Freedom of choice: Ensuring that migrants can opt for a migration assistance process without physical or psychological pressure. b) Informed decision: Guaranteeing that individuals seeking assistance have access to updated, impartial, and reliable information to make an informed decision about their return.

    \textbf{a) Freedom of choice:} 
    This is defined when there are no physical or psychological pressures for the migrant to undergo a migration assistance process.

    \textbf{b) Existence of an informed decision:} 
 
This point ensures that the person eligible for assistance has all the relevant information, and that it is up-to-date, impartial, and reliable. In this way, the migrant's decision is ensured.

\section{Situations of Return Migration:}

In recent years, there has been an unconsciously witnessed massive return migration, driven by the need for economic survival. In Peru, this was evident with the arrival of COVID-19 in 2020, as people sought this resource to sustain themselves economically.

\subsection{Return Migration in the Context of a Pandemic:}

Rufino (2020) states that when the Peruvian government declared a nationwide health emergency, more than 165,000 Peruvians attempted to return to their places of origin. Hundreds of people who had migrated from provinces to major cities returned to their communities when their idealized urban lives collapsed due to economic activities coming to a halt, leaving them unemployed and at high risk of virus contagion.
A press release from the Ministry of Transportation and Communications (MTC, 2020) indicates that until May 2020, 380 humanitarian flights were authorized for Peruvians and foreigners stranded due to the closure of Peruvian borders.
Upon returning to their communities, people began working in activities typical of their areas, such as agriculture, livestock, fishing, among others, aiming to achieve subsistence. Being unemployed and lacking economic income, any means of resource acquisition was welcome.
With return migration, even if not for an extended period, populated centers, communities, and settlements were once again inhabited. Previous activities resumed, allowing people to live by leveraging the resources available in their localities.

\section{MATERIALS AND METHODS:}

The research had a mixed-methods approach. In the quantitative focus, measurements obtained from surveys were analyzed, while in the qualitative aspect, norms were evaluated to develop normative proposals that could be included in the Statute of the Antacahua Community. The research is non-experimental, as variables were not manipulated or controlled. It is cross-sectional because data collection occurred at a single point in time. The research's scope is descriptive, seeking to describe the characteristics of the causes that lead to community members' migration and proposing alternatives to mitigate this migration.
A case study was conducted with a population of 96 community members from the Antacahua Peasant Community, established 34 years ago, located in the Huancané province of the Puno department at an altitude of 3855 meters above sea level. The survey sample conducted on November 14, 2022, consisted of 64 community members, of which 45.3\% were male and 54.7\% female. Additionally, 50\% of them were qualified community members, 45.3\% were natives, and only 4.7\% were integrated community members. Concerning their education, 87.5\% had incomplete primary education, 10.9\% completed primary education, and only 1.6\% completed secondary education. Data analysis was performed in an Excel spreadsheet, generating bar graphs subsequently interpreted. Bibliographical and legal analysis was conducted to propose new clauses or precepts to the Community Statute. The results could serve as an example for communities with similar productive activities and/or similar geographical, social, political, and economic characteristics.

\section{RESULTS AND DISCUSSION:}

Figure 1, derived from the survey conducted in the Community Campesina de Antacahua and the migrant population, shows that:

32\% migrate due to basic needs, seeking housing with complete basic services.
23\% migrate due to lack of economic support.
17\% do so because of the low productivity of crops.
19\% of the migrant community migrates for study-related reasons.
9\% migrate for other reasons, including the lack of land.
[Figure 1 - Migration Motives] (A description of the figure or its content can be provided if available).

\textbf{Figura 1.}

\vspace{0.5cm}
   \includegraphics[width=0.4\textwidth]{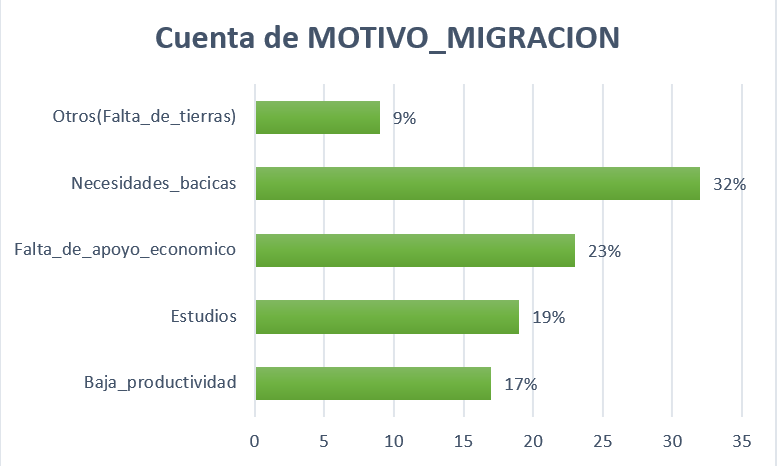}

In Figure 2, derived from the survey conducted in the Community Campesina de Antacahua and the migrant population, according to age groups, it can be observed that:

14\% of migrants are teenagers between 15 and 22 years old.
22\% are young adults aged 23 to 29.
16\% are adult-youth migrants aged 30 to 36.
48\% are adult migrants aged 37 to 60.
[Figure 2 - Age Group of Migrants] (A description of the figure or its content can be provided if available).

\textbf{Figura 2.}

\vspace{0.5cm}
   \includegraphics[width=0.4\textwidth]{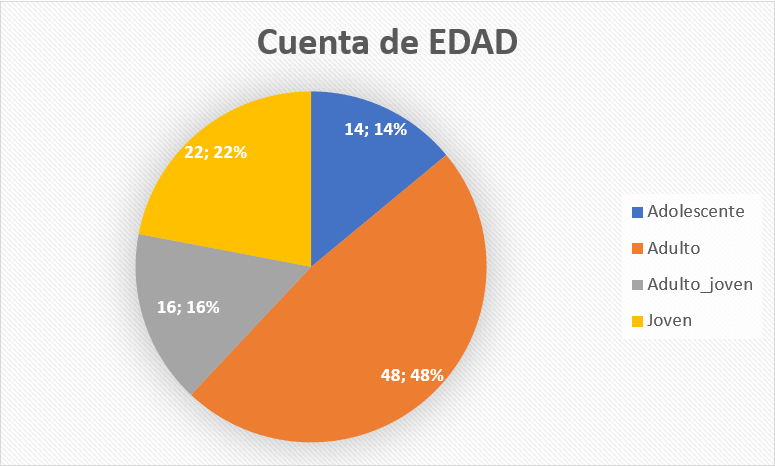}
   
Figure 3, obtained from the survey conducted in the Community Campesina de Antacahua and the migrant population, according to the most frequent places to migrate, shows that:

40.0\% migrate to the city of Juliaca, which is a predominantly commercial city.
22.0\% migrate to the city of Puno.
25.0\%, being the second most preferred place to migrate, consists of the mines located in Ananea and Cojata.
13.0\% migrate to other places, apart from those mentioned.
[Figure 3 - Most Frequent Places to Migrate Temporarily or Permanently] (A description of the figure or its content can be provided if available).

\textbf{Figura 3.}

\vspace{0.5cm}
   \includegraphics[width=0.4\textwidth]{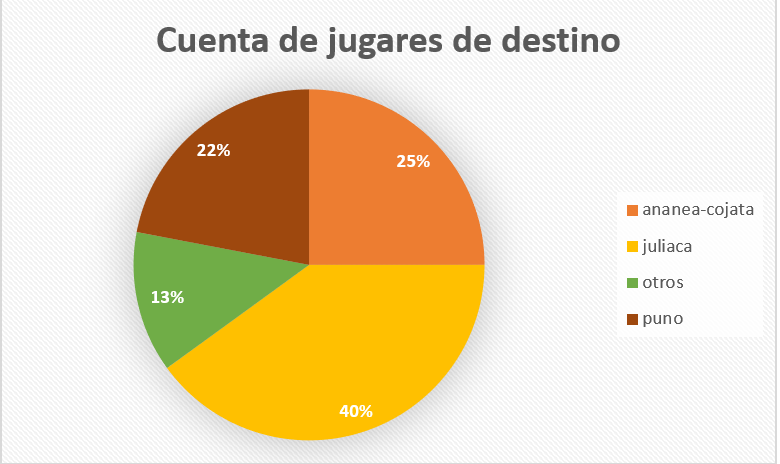}
   
The results presented in Figures 1, 2, and 3 indicate that the population migrates due to the lack of services that satisfy basic needs, economic scarcity, and a lack of opportunities. This aligns with a study conducted by Enrique Rodríguez (1999), who points out that living conditions in rural areas are very precarious. This situation is primarily due to limited resources resulting from strong population pressure on land, as evidenced by indicators such as land ownership, possession of livestock, income, and overall capital availability. However, an important contributing factor to this situation is also the deficient state observed in productive and social infrastructure.
The age group most prone to migration is adults aged 30 to 36, and the preferred destination is the city of Juliaca. This is attributed to the opportunities presented in a highly commercial city, along with greater prospects for employment and economic generation, making it attractive for a population that lacks support.
In the results of a study by Jorge Carbajal and Hanna Serrano (2022) conducted in the Cusco Department between 2007 and 2017, rural-urban migration is explained by push factors such as unemployment and inefficient access to basic services. To make migration from rural to urban areas attractive to residents, there is a greater promotion of employment, implying perceived income and economic and social infrastructure. This confirms the hypothesis that rural-urban migration is explained by economic income, employment, and access to services, indicating a strong correlation between the migration variable and income, employment, and access to basic services. The research by Carbajal and Serrano demonstrates that rural-urban migration for the purpose of seeking a better quality of life economically and socially is not limited to the Antacahua Peasant Community in the Puno Region but is also observed in other regions such as Cusco.
According to María Tona (2021), the migratory process in Argentina is closely linked to the fluctuations in national economic development. Its impact is evident in the loss or increase of population, with the province of Buenos Aires experiencing the most noticeable population imbalances. Cases are evident where groups of people decide to leave their rural homes and move to urban centers, indicating that rural-urban migration in Buenos Aires reveals structural conditions and uprooting to seek a "better life" in the big city.
This research suggests that, similar to some regions in Peru, in Argentina, there is also migration of rural groups to urban areas in search of better opportunities and a higher quality of life. These factors are common to the situation observed in the Antacahua Peasant Community.
\section{}
\vspace{0.5cm}
   \includegraphics[width=0.4\textwidth]{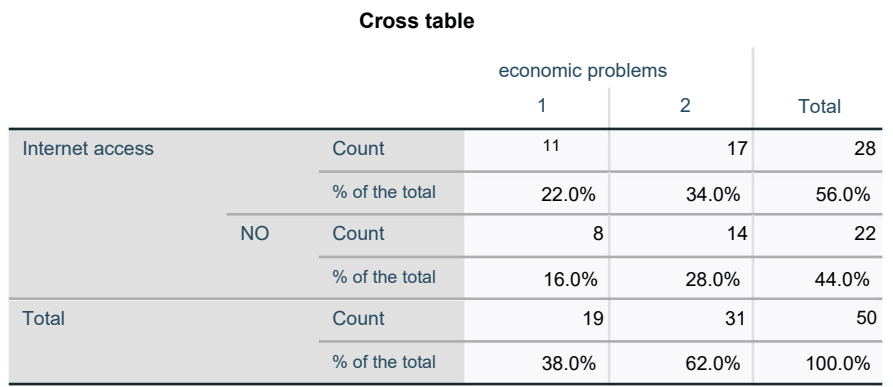}
   \section
   Out of a total of 100\% of the respondents, with 50 individuals surveyed, 22\% have economic problems but have internet, 16\% have economic problems but do not have internet, 34\% have no economic problems and have internet, and 28\% have economic problems but do not have internet. In total, 56\% have internet, and 44\% do not have internet.
\vspace{0.5cm}
   \includegraphics[width=0.4\textwidth]{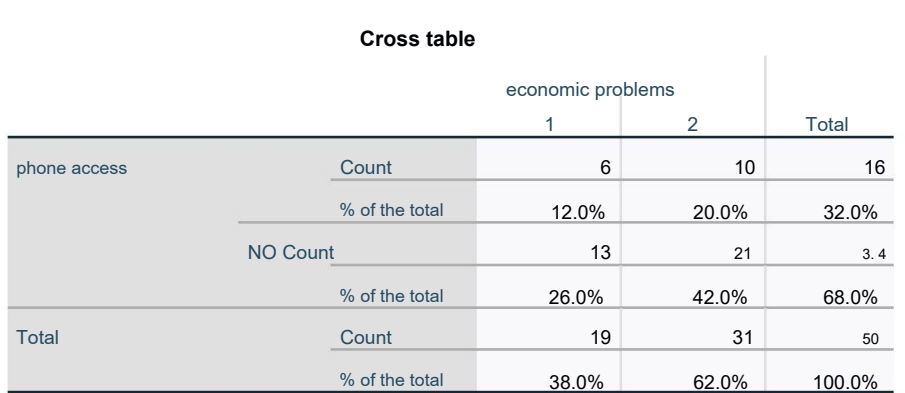}
   Out of a total of 100\% of the respondents, with 50 individuals surveyed, 12\% have access to a phone with economic problems, 26\% do not have access to a phone with economic problems, 20\% have access to a phone with economic problems, and 42\% do not have access to a phone without economic problems. In total, 32\% of respondents have access to a phone, and 68\% of respondents do not have access to a phone.

\vspace{0.5cm}
   \includegraphics[width=0.4\textwidth]{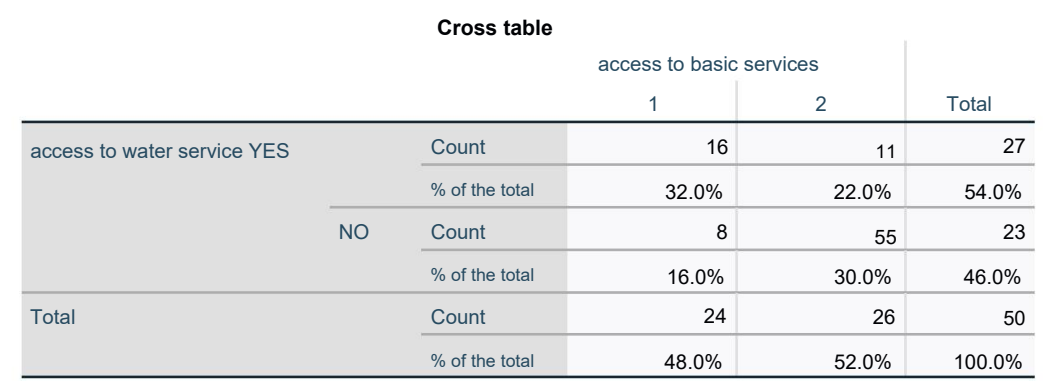}
   Out of a total of 100\% of respondents, with 50 individuals surveyed, 32\% have both basic services and water service, 16\% do not have water service but have basic services, 22\% have water service but do not have basic services, 30\% have neither water service nor basic services. In total, 54\% have water service, and 46
   
\vspace{0.5cm}
   \includegraphics[width=0.4\textwidth]{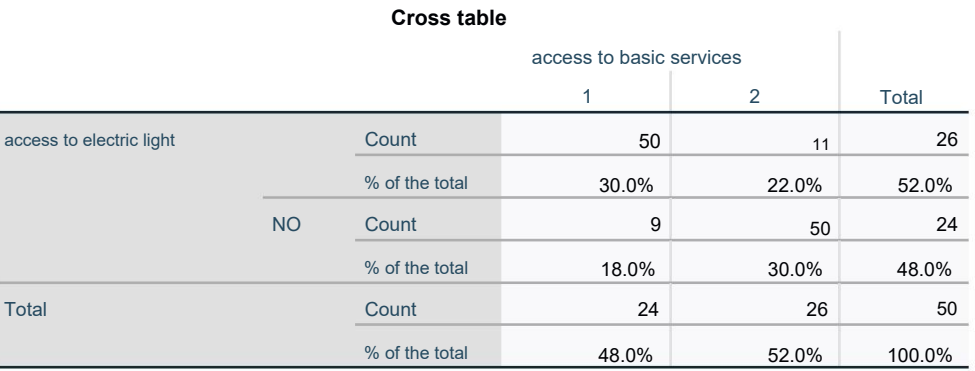}
   Out of a total of 100\% of respondents, with 50 individuals surveyed, 30\% have electrical service and basic services, 18\% do not have electrical service but have basic services, 22\% have electrical service but not basic services, 30\% have neither electrical service nor basic services. In total, 52\% out of 100\% have electrical service, and 48\% out of 100\% do not have electrical service.
   
\vspace{0.5cm}
   \includegraphics[width=0.4\textwidth]{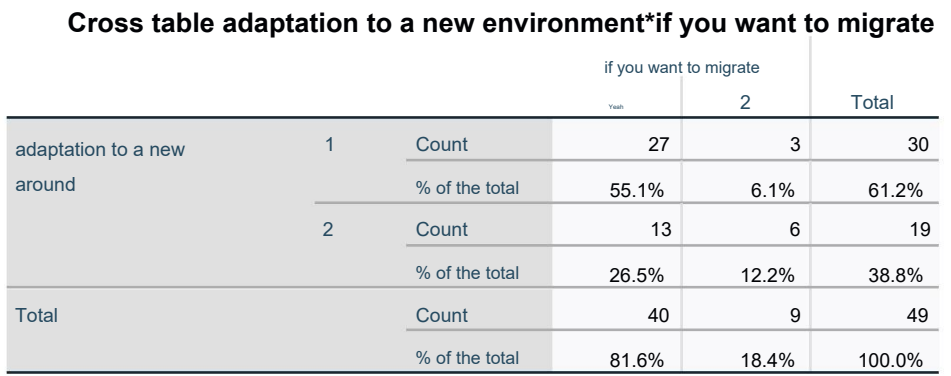}
   Out of a total of 100\% of respondents, with 50 individuals surveyed, 55.1\% wanted to migrate and adapted to their new life, 26.5\% migrated but did not adapt to their new life, 6.1\% did not migrate and are adapted to their life, and 12.2\% did not adapt to their life and did not migrate. In total, 81.6\% of surveyed individuals migrated, and 18.4\% did not migrate. 61.2\% of individuals feel comfortable in their life, while 38.8\% do not feel comfortable or adapted in their life.
\vspace{0.5cm}
   \includegraphics[width=0.4\textwidth]{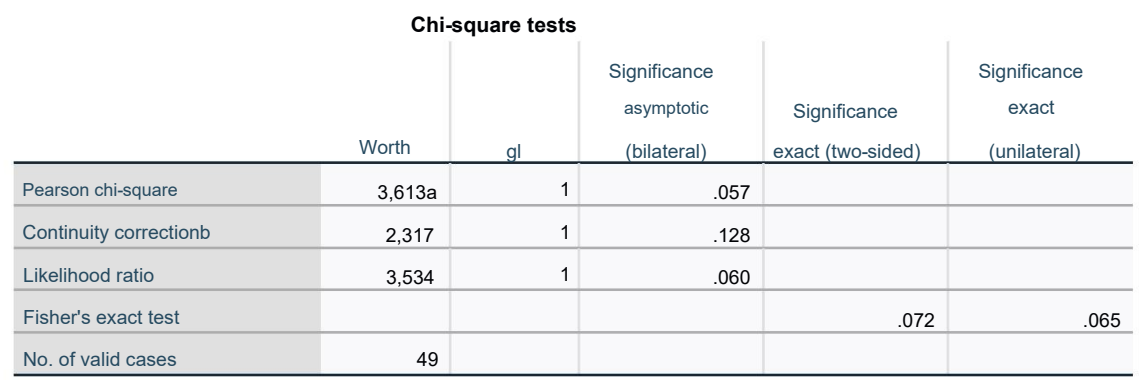}
   According to the statistical method of chi-square, the null hypothesis is not rejected, and the alternative hypothesis is accepted.

\section{POLICIES TO MITIGATE MIGRATION IN THE ANTACAHUA PEASANT COMMUNITY}

At the national level, various public policies have been implemented to modify migratory flows. These include initiatives such as affirming national identity, eradicating violence, strengthening civic values and citizen security, political, economic, and administrative decentralization to promote development, affirmation of the social market economy, agrarian and rural development policies, poverty reduction efforts, universal access to free education, public health initiatives, eradication of coca production, and the elimination of terrorism.
In the Antacahua Peasant Community, two types of migration have been identified: temporary and permanent migrations, both driven by different causes.
Firstly, considering the reasons for migration, the lack of basic services represents the highest percentage at 34.4\%. This implies that many community members migrate due to the absence of essential services such as water, electricity, internet, telephone, and education. An interesting policy proposal within the framework of Article 89 of the Constitution of Peru, which states that "Peasant Communities (...) are autonomous in (...) the free disposition of their lands (...)," is the introduction of the following provision to the Community Statute: "If there is a public or private institution interested in providing services within the community, the community donates, with the possibility of reversion, a necessary portion of land for its installation."
Secondly, to reduce emigration from the Antacahua Community, it is crucial to improve the community's economic conditions. This involves promoting employability by enhancing the knowledge, skills, and competencies of individuals based on the needs of the labor market. An interesting alternative is to encourage and train community members to provide agro-veterinary assistance services. Proposed provisions for the Statute include, "As an additional member of the Communal Board, a Secretary of Productive Activities is elected; this person must have accredited technical education at a minimum, and the Community is obligated to remunerate a salary equivalent to 50\% of a UIT," and functions such as facilitating the transformation of eucalyptus wood, seeking agreements for community members to exploit wood more profitably, and promoting adequate and quality training in agro-veterinary matters.
Thirdly, based on the Law of Peasant Communities, which allows the constitution of communal and multicommunal enterprises, a proposal is made to engage community families in communal economic activities. Specifically, there is a suggestion to promote communal breeding of high-quality genetic guinea pigs as a potential source of income for families. The proposal involves the election of a Secretary of Productive Activities, as well as functions related to the formation of communal enterprises, seeking agreements for wood exploitation, and facilitating the growth of agro-veterinary services.
In conclusion, these proposed policies aim to address the root causes of migration in the Antacahua Peasant Community by improving access to essential services, enhancing economic opportunities, and fostering communal economic activities.

\section{CONCLUSION}

Migration is a social phenomenon where a group of people leaves their place of residence to settle elsewhere, typically from rural to urban areas, seeking better economic and social conditions.
According to the research conducted in the Antacahua Community, migration is primarily attributed to the lack of basic services, such as access to electricity, drinking water services, internet, etc. It is also driven by a lack of economic support, particularly in terms of providing land for agricultural and livestock productivity. The age group that migrates the most in the community comprises young adults between the ages of 30 and 36, and the most common destination is the city of Juliaca, which is economically active.
Comparing the situation in the Antacahua Peasant Community at the national and international levels, we find that in the Cusco Province in Peru, rural-urban migration also occurs due to economic and social reasons. The rural population in the districts of Cusco migrates in search of a better quality of life. On an international scale, a similar situation is observed in Argentina, where the rural population of Buenos Aires feels disadvantaged compared to urban areas and, therefore, seeks a better quality of life by migrating to those destinations. According to research, the main causes of rural-urban migration are primarily the lack of economic support, low productivity, and significant social disparities.
To address migration from a normative perspective expressed in the Community Statute, provisions have been proposed to facilitate the installation of basic services, provide land to service-providing businesses, and promote economic activity. This includes adding a member to the Communal Board responsible for community entrepreneurial initiatives, particularly in areas such as agro-entrepreneurship, eucalyptus transformation, and guinea pig breeding.

\newpage
\section{BIBLIOGRAPHICAL REFERENCES}

\begin{itemize}
    \item Aruj, R. (2008). Causas, consecuencias, efectos e impacto de las migraciones en Latinoamérica. \textit{Papeles Poblac. 2008;(55):22}. Recuperado de \url{https://www.scielo.org.mx/pdf/pp/v14n55/v14n55a5.pdf}
\end{itemize}

\begin{itemize}
    \item Carbajal, J., \& Serrano, H. (2022). ``Migración Rural-Urbana de Cusco y su Impacto en el Crecimiento Económico de la Provincia de Cusco, 2007-2017'' [Tesis para optar título profesional]. Universidad Andina del Cusco.
\end{itemize}

\begin{itemize}
    \item Espejo, A. (2017). Inserción laboral de los jóvenes rurales en América Latina. Un breve análisis descriptivo. \textit{Grupos de Diálogo Rural, una estrategia de incidencia.} Recuperado de \url{https://www.rimisp.org/wpcontent/files_mf/1502548172Inserci%C3%B3nlaboraldelosj%C3%B3venesruralesenAm%C3%A9ricaLatina.pdf}
\end{itemize}

\begin{itemize}
    \item Guillén, J. C., Menéndez, F. G., y Moreira, T. K. (2019). Migración: Como fenómeno social vulnerable y salvaguarda de los derechos humanos. \textit{Revista de Ciencias Sociales (Ve), XXV(E-1), 281-294.} doi: \url{10.31876/rcs.v25i1.29619}
\end{itemize}

\begin{itemize}
    \item Gutiérrez Silva, J. M., Romero Borré, J., Arias Montero, S. R. y Briones Mendoza, X. F. (2020). Migración: Contexto, impacto y desafío. Una reflexión teórica. \textit{Revista de Ciencias Sociales (Ve), XXVI(2), 299-311.} Recuperado de \url{https://www.redalyc.org/articulo.oa?}
\end{itemize}

\begin{itemize}
    \item Instituto Nacional de Estadística e Informática [INEI] (2011). Perú: Perfil de la pobreza por departamentos, (2001-2010). Recuperado de \url{https://www.inei.gob.pe/media/MenuRecursivo/publicaciones_digitales/Est/Lib0801/libro.pdf}
\end{itemize}

\begin{itemize}
    \item Instituto Nacional de Estadística e Informática (2015). Migraciones Internas en el Perú a nivel departamental. \textit{Organización Internacional para las Migraciones – Misión en el Perú.} Recuperado de \url{https://peru.iom.int/sites/g/files/tmzbdl951/files/Documentos/20-03-2017_Publicaci%C3%B3n%20Migracion%20Interna%20por%20Departamentos%202015_OIM.pdf}
\end{itemize}

\begin{itemize}
    \item Instituto Nacional de Estadística e Informática. (2020) Efectos de la migración Interna sobre el crecimiento y la estructura demográfica 2012-2017. Recuperado de \url{https://www.inei.gob.pe/media/MenuRecursivo/publicaciones_digitales/Est/Lib1732/libro.pdf}
\end{itemize}

\begin{itemize}
    \item Lazos Chavero, E. (2020). Advierten sobre amenazas al campo mexicano. \textit{Boletín UNAM-DGCS-597. DGCS.} Recuperado de \url{https://www.dgcs.unam.mx/boletin/bdboletin/2020_597.html}
\end{itemize}

\begin{itemize}
    \item León, L. A. (2015). Análisis económico de la población. Demografía. \textit{Departamento Académico de Economía de la FACEAC de la Universidad Nacional “Pedro Ruiz Gallo” de Lambayeque. Perú.} Recuperado de \url{https://web.u a.es/es/giecryal/documentos/demografia-peru.pdf}
\end{itemize}

\begin{itemize}
    \item Organización Internacional de las Migraciones [OIM]. (2022). \textit{Glosario de la OIM sobre Migración.} Recuperado de \url{https://publications.iom.int/es/system/files/pdf/iml-34-glossary-es.pdf}
\end{itemize}

\begin{itemize}
    \item Ministerio de Transportes y Comunicaciones [MTC]. (2020) \textit{MTC autorizó 380 vuelos internacionales humanitarios en el marco de la emergencia sanitaria.} Recuperado de \url{https://www.gob.pe/institucion/mtc/noticias/143647-mtc-autorizo-380-vuelos-internacionales-humanitarios-en-el-marco-de-la-emergencia-sanitaria}
\end{itemize}

\begin{flushleft}
    Portal de Datos Mundiales sobre Migración. (2022). Migración de retorno. \url{https://www.migrationdataportal.org/es/themes/return-migration}
\end{flushleft}

\begin{flushleft}
    Rodríguez, E. (1999). Entre el campo y la ciudad: estrategias migratorias frente a la crisis. En \textit{Estrategias de supervivencia y seguridad alimentaria en América Latina y en África.} (Sur, sur, Vol. 1, p. 71). CLACSO, Consejo Latinoamericano de Ciencias Sociales. Recuperado de \url{http://bibliotecavirtual.clacso.org.ar/clacso/sur-sur/20100707020524/5_doig.pdf}
\end{flushleft}

\begin{flushleft}
    Rufino. M. A. (2020). \textit{COVID19 y la migración de retorno en el norte de Perú.} CAS. Recuperado de \url{https://calacs.wp.st-andrews.ac.uk/regresando-a-mis-raices-covid19-y-la-migracion-de-retorno-en-el-norte-de-peru/}
\end{flushleft}

\begin{flushleft}
    Silva del Carpio, M. A. (2018). \textit{Caracterización de las nuevas empresas “comunales”. El caso de las empresas comunales de la comunidad campesina la encañada en Cajamarca.} Fondo Editorial PUCP. Recuperado de \url{https://tesis.pucp.edu.pe/repositorio/bitstream/handle/20.500.12404/12694/SILVA_CARPIO_MARIA_CARACTERIZACION_NUEVAS_EMPRESAS.pdf?sequence=4&isAllowed=y}
\end{flushleft}

\begin{flushleft}
    Tona, M. (2021). Vista de \textit{Memorias de la migración rural en clave de género (Buenos Aires, a partir de mediados del siglo XX) | Testimonios.} Revistas de la Universidad Nacional de Córdoba. Recuperado de \url{https://revistas.unc.edu.ar/index.php/testimonios/article/view/36330/36664}
\end{flushleft}

\begin{flushleft}
    Weller, J. (2006). \textit{Los jóvenes y el empleo en América Latina. Desafíos y perspectivas ante el nuevo escenario laboral.} Comisión Económica para América Latina y el Caribe. Pág. 184-190. Recuperado de: \url{https://repositorio.cepal.org/bitstream/handle/11362/1902/S33134W448_es.pdf}
\end{flushleft}

\end{document}